\documentclass[aps,prl,twocolumn,superscriptaddress,amsmath,amssymb,floats]{revtex4}

\usepackage{graphicx}
\usepackage{dcolumn}
\usepackage{bm}
\usepackage{amsmath}
\usepackage{amssymb}
\usepackage{epsfig}
\usepackage{array}
\usepackage{times}
\begin{document} \title{Optical spectra of the heavy fermion uniaxial ferromagnet UGe$_2$}

\author{V.\ Guritanu}
\affiliation {D\'epartement de Physique de la Mati\`ere Condens\'ee, Universit\'e de Gen\`eve, quai Ernest-Ansermet 24, CH 1211, Gen\`eve 4, Switzerland}

\author{N.\ P.\ Armitage}
\affiliation{Department of Physics and Astronomy, The Johns
Hopkins University, Baltimore, MD 21218, USA}

\author{R.\ Tediosi}
\affiliation {D\'epartement de Physique de la Mati\`ere Condens\'ee, Universit\'e de Gen\`eve, quai Ernest-Ansermet 24, CH 1211, Gen\`eve 4, Switzerland}

\author{S.\ S.\ Saxena}
\affiliation{Department of Physics, Cavendish Laboratory,
University of Cambridge, Madingley Road, Cambridge CB3 0HE, UK}

\author{A.\ Huxley} 
\affiliation{Scottish Universities Physics Alliance, School of Physics,
King's Buildings, University of Edinburgh, Edinburgh EH9 3JZ, UK}

\author{D.\ van der Marel}
\affiliation {D\'epartement de Physique de la Mati\`ere Condens\'ee, Universit\'e de Gen\`eve, quai Ernest-Ansermet 24, CH 1211, Gen\`eve 4, Switzerland}

\date{\today }

\begin{abstract}

We report a detailed study of UGe$_{2}$ single crystals using infrared reflectivity and spectroscopic
ellipsometry. The optical conductivity suggests the presence of a low frequency interband transition and a narrow free-carrier response with strong frequency dependence of the scattering rate and effective mass. We observe sharp changes in the low frequency mass and scattering rate below the upper ferromagnetic transition $T_C = 53 K$. The characteristic changes are exhibited most strongly at an energy scale of around 12 meV (100 cm$^{-1}$).  They recover their unrenormalized value above $T_C$ and for $\omega >$ 40 meV.  In contrast no sign of an anomaly is seen at the lower transition temperature of unknown nature $T_x \sim$ 30 K, observed in transport and thermodynamic experiments.  In the ferromagnetic state we find signatures of a strong coupling to the longitudinal magnetic excitations that have been proposed to mediate unconventional superconductivity in this compound.

\end{abstract}
\maketitle

The possibility of unconventional superconductivity mediated by ferromagnetic fluctuations has long been a subject of theoretical speculation \cite{ginzburg1957,fay1980}. Interest in this subject has been recently piqued with the discovery of superconductivity coexisting with the ferromagnetic state of UGe$_{2}$ under pressure \cite{saxena2000}. UGe$_{2}$ is a strongly anisotropic uniaxial ferromagnet with filled $5f$ electron states. Due to correlations and conduction band-$5f$ hybridization, carrier masses are found to be strongly enhanced \cite{onuki1991} (10 - 25 m$_{0}$) although specific heat coefficients still fall an order of magnitude short of the largest values found in antiferromagnetic uranium-based heavy fermion (HF) compounds. UGe$_{2}$ exhibits a Curie temperature that strongly decreases with increasing pressure from about 53 K at ambient pressure to full suppression around 16 kbar.  Superconductivity exists in a pressure region from 10 to 16 kbar, just below the complete suppression of ferromagnetism  \cite{saxena2000}.  Although superconductivity and ferromagnetism are usually found to be antagonistic phenomena, the observation fits within the now common scenario of finding superconductivity near the zero temperature termination of a magnetic phase. In this sense it seemed quite natural to associate the superconductivity with being mediated by the magnetic fluctuations that diverge at a quantum critical point (QCP) perhaps as in the case of pressure driven superconductivity in the antiferromagnetic HF superconductors. However, the para- to ferromagnetic transition is strongly first-order and is not associated with a peak in the effective electronic mass or superconducting transition temperature. It therefore appears that superconductivity is not directly related to the quantum phase transition connecting the ferromagnetic and paramagnetic states \cite{huxley2001,pfleiderer2002}. 

In addition to the main ferromagnetic transition,  there appears to be an additional weak first-order transition  of more enigmatic origin at lower temperatures (for $p$ = 0; $T_{x}$ = 30 K and for $T$ = 0; $p_x \approx$ 12.5 kbar), which has been identified $via$ resistivity  \cite{huxley2001,oomi1998,tateiwa2001}, magnetization \cite{pfleiderer2002} and heat capacity \cite{oomi1998,tateiwa2001}. The critical pressure $p_x$ is very close to that where the superconducting temperature is maximum, and it may be that the superconductivity is mediated $via$ fluctuations from this weak first-order transition. The exact nature of the state below $T_x$ is not clear at present. It has been suggested that these two magnetic phases are related by a first-order Stoner-like phase transition in the spin magnetization due to a sharp double-peak in the density of states near $E_F$ \cite{sandeman2003}. An alternative suggestion is that a competition between spin-orbit coupling and crystal field effects drives a change in the local moment configuration \cite{shick2004}. Other possibilities such as a coexisting charge-density wave or spin-density wave order exist, however neutron scattering has failed to detect any such phases thus far \cite{huxley2001}. We also note that at ambient pressure the signatures of the lower transition $T_x$ is only weakly visible in the resistivity \cite{oomi1995} and magnetization measurements \cite{pfleiderer2002,motoyama2001}. Moreover, there is also no or almost no anomaly in the specific heat at ambient pressure \cite{huxley2001,tateiwa2001}. It may be that this transition at $T_x$ does not extend all the way to zero pressure and, in fact, the line of first order transitions terminates in a critical end point at finite temperature and pressure.  $T_x$ at ambient pressure is then indicative of a a crossover and not a true phase transition.

In this paper we present the results of a detailed optical study of UGe$_2$ single crystal using infrared reflectivity and spectroscopic ellipsometry. We have found a renormalized zero frequency mode with a large frequency dependent effective mass and scattering rate below the upper ferromagnetic transition $T_C$. They recover their unrenormalized values above $T_C$ and for $\omega >$ 40 meV. In contrast no sign of an anomaly is seen at $T_x \sim$ 30 K. In the ferromagnetic state, we find signatures of a strong coupling to the longitudinal magnetic fluctuations which have been proposed to mediate unconventional superconductivity in this compound.

Measurements were performed in the frequency range from 50 cm$^{-1}$ (6.2 meV) to 30000 cm$^{-1}$ (3.7 eV) combining infrared (IR) reflectivity via Fourier transform spectroscopy, and ellipsometry from the near-infrared to ultraviolet energy range. The absolute value of the reflectivity $R(\omega,T)$ was calibrated against a reference gold layer evaporated \textit{in situ} on the sample surface. Ellipsometry and IR data were combined using a Kramers-Kronig consistent variational fitting procedure \cite{kuzmenko2005}. This allows the extraction over a broad energy range of all the significant frequency and temperature dependent optical properties like for instance, the complex conductivity $\sigma(\omega,T)= \sigma_1 + i\sigma_2$. The sample used for this study was grown at the  CEA-Grenoble by the Czochralski technique \cite{saxena2000}. Measurements were taken in quasi-normal incidence to the a-c plane using linearly polarized light. We found only a small shift of the spectra between the two crystal directions with any differences smaller than our experimental accuracy ($1\%$). The displayed spectrum is the average of the two directions.  
\begin{figure}
	\centering
		\includegraphics{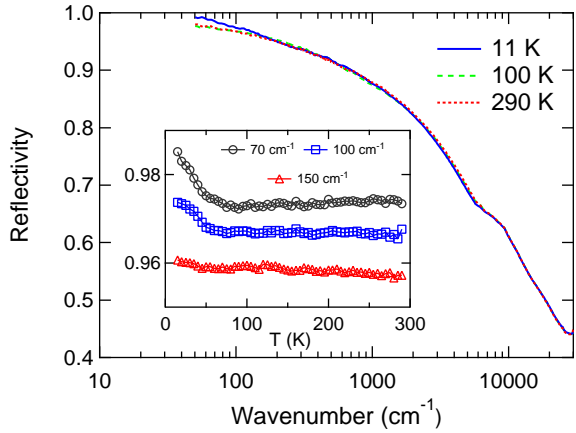}
	\caption{(Color online) The normal-incidence reflectivity $R(\omega)$ of UGe$_2$ at three temperatures. The inset shows $R(\omega)$ at 70, 100 and 150 cm$^{-1}$. The reflectivity curves at 100 K and 290 K overlap on this scale.}
	\label{r}
\end{figure}

Figure \ref{r} shows the reflectivity spectra $R(\omega)$ over the full energy range.  At high temperature $R(\omega)$ exhibits a monotonic increase as expected for a metal with $R(\omega)\rightarrow1$ as $\omega\rightarrow 0$.  As the sample is cooled below $T_C$ the reflectivity shows a significant increase with the largest effects at low frequency (see Fig. \ref{r} inset). No sign of an anomaly is found at the temperature $T_x$, which may be not surprising considering its weak signature in transport and thermodynamics at ambient pressure. It is interesting to mention that URu$_2$Si$_2$ also has a second magnetic transition of unknown nature at low temperatures, which appears very pronounced in the infrared spectrum \cite{bonn1988}. The fact that in UGe$_2$ we see no signature of the second magnetic transition in the optical spectra, suggests a very different nature of this transition in the two materials.

\begin{figure}
	\centering
		\includegraphics{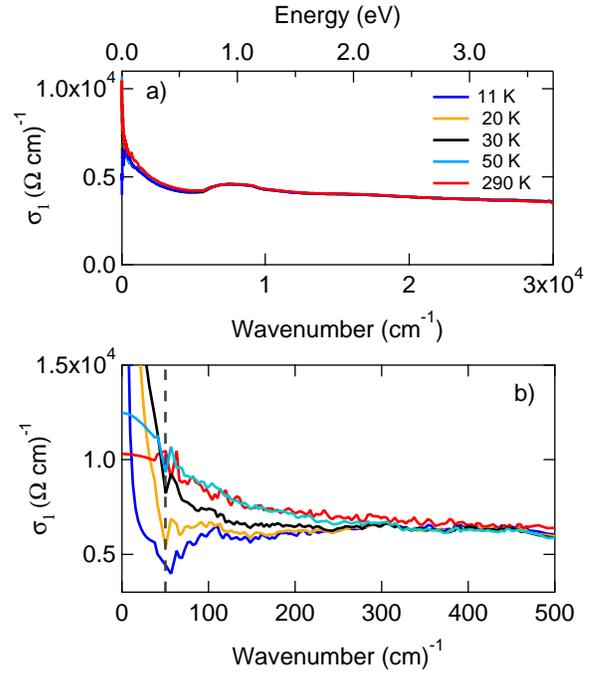}
	\caption{(Color) a) $\sigma_1(\omega)$ at different temperatures in the energy range of 6.2 meV to 3.7 eV. b) The enlargement of the low energy data including the zero frequency extrapolations.} 
	\label{s1}
\end{figure}

We observe that the reflectivity has no significant temperature dependence above 1000 cm$^{-1}$ (0.12 eV), which suggests a negligible temperature dependence at even higher frequencies. We used therefore room temperature ellipsometry for frequencies greater than 6000 cm$^{-1}$ (0.74 eV) to calculate the complex optical conductivity over the entire spectral range displayed in Fig. \ref{s1} a). We assign the peak at approximately 1 eV to an interband transition. At high temperature the low frequency optical conductivity $\sigma_1(\omega)$ shows a broad Drude-like behaviour as expected for a metal. The low frequency conductivity is strongly temperature and frequency dependent. To clarify this trend we show an enlargement of the low frequency data in Fig. \ref{s1} b). Below 50 cm$^{-1}$ we plot the zero frequency extrapolations obtained using the DC conductivity in the fitting procedure. Several remarkable low frequency structures form at low temperatures out of the broad Drude peak, including a very narrow zero frequency mode which represents the intraband response of the heavy quasiparticles.  Concomitantly, a maximum develops at 110 cm$^{-1}$ (13.6 meV) and a weak structure at 300 cm$^{-1}$ (37 meV). As will be shown below, these features reflect various aspects of the coherent HF and ferromagnetic states.

In order to further analyze the shape of the low frequency spectra we use the extended Drude model. In this formalism the optical constants are expressed in terms of  a frequency dependent effective mass $m^{\ast}(\omega)/m$ and scattering rate 1/$\tau(\omega)$ by the following expression

\begin{eqnarray}
\frac{m^{\ast}(\omega)}{m}= - \frac{\omega^{2}_{p}}{4\pi\omega}Im\left[\frac{1}{\sigma(\omega)}\right],
 \hspace{1.5mm}
\frac{1}{\tau(\omega)}= \frac{\omega^{2}_{p}}{4\pi}Re\left[\frac{1}{\sigma(\omega)}\right]
\label{mstar}
\end{eqnarray}

where $\hbar\omega_{p} = \sqrt{4\pi n e^2/m_b}=3.5$ eV is the total Drude plasma frequency and $\sigma(\omega)$ is the complex optical conductivity.  $\hbar\omega_{p}$, which is determined through $m^{\ast}(\omega)/m = 1$ at 290 K and $\omega$ = 500 cm$^{-1}$, acts as a normalization constant and does not affect the trends as a function of $\omega$ and $T$ .

\begin{figure}
	\centering
		\includegraphics{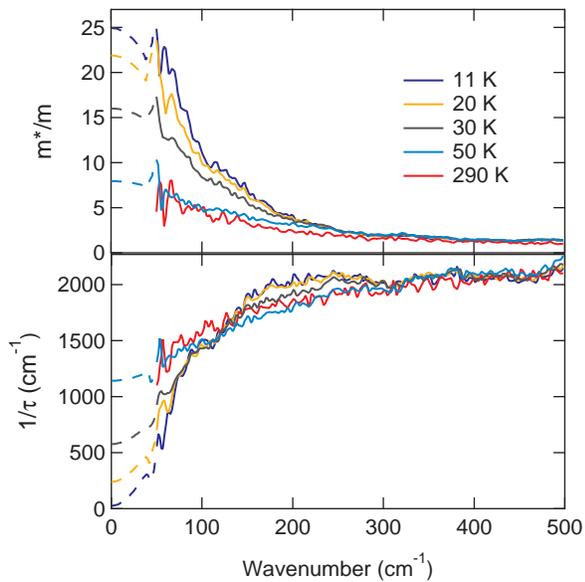}
	\caption{(Color) The effective mass and the scattering rate as a function of photon energy derived from the extended Drude model with $\hbar\omega_{p} = 3.5$ eV. The dashed lines below 50 cm$^{-1}$ show the extrapolation toward zero frequency obtained using the DC transport in the fitting process.}
	\label{m_tau}
\end{figure}

Figure \ref{m_tau} displays the spectra of $1/\tau(\omega)$ and $m^*(\omega)/m$ as a function of frequency obtained from the extended Drude model for different temperatures. At room temperature both the scattering rate and the effective mass are nearly frequency independent.  As the sample is cooled and the magnetic order develops the effective mass is strongly enhanced and the scattering rate suppressed.  Such behaviour suggests the development of heavy quasiparticles at low temperature. We observe that the strong increase of the effective mass and the rapid decrease of the scattering rate are largest below approximately 100 cm$^{-1}$ (12 meV) which can be regarded as the characteristic energy of the heavy quasiparticles. Below 50 cm$^{-1}$ we use the extrapolation toward zero frequency which gives a quasiparticle effective mass of over 25 for $\omega\rightarrow 0$ at the lowest $T$. Qualitatively this value is consistent with the cyclotron masses of 10 - 25 m$_0$ obtained by dHvA measurements \cite{onuki1991}.

\begin{figure}
	\centering
		\includegraphics{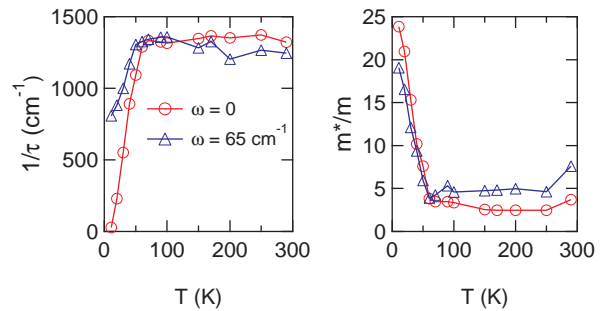}
	\caption{(Color online) Temperature dependence of the effective mass and the scattering rate for two frequencies.}
	\label{m_tau_temp}
\end{figure}

In Figure \ref{m_tau_temp} we present the temperature dependence of the effective mass and scattering rate at 0 and 65 cm$^{-1}$ obtained from Fig. \ref{m_tau}. Note that at high temperature, in the paramagnetic state, both quantities are temperature independent. However, starting exactly at $T_C$ the scattering rate becomes strongly suppressed and the effective mass enhanced.  Such behavior contrasts with the usual situation in HF compounds, where mass renormalizations develop below a coherence temperature $T^*$ different than the transition temperature to a magnetic state. Moreover, the conventional view is that the coherent HF state should actually be $suppressed$ at a magnetic transition as the dominance of the RKKY interaction is expected to quench the Kondo effect \cite{doniach1977}.  The observed behavior is not completely unprecedented however, as a number of other multi-$f$ electron compounds have been found to undergo additional mass enhancements at the magnetic transition \cite{dressel2002}.  In the present case however the magnetic transition appears to actually trigger the HF state suggesting an intrinsic $T^* < T_C$. This behavior may be related to multiple occupation of the 5$f$ levels, although we note that a similar effect in 3$d$ electron systems has also been observed at the $T_C$ of ferromagnetic Yb$_{14}$MnSb$_{11}$ \cite{burch2005} and helimagnetic FeGe \cite{guritanu2007}.

In the optical spectra shown in Fig. \ref{m_tau} we observe a rather strong, but incomplete, suppression of $1/\tau(\omega)$ for frequencies smaller than about 50 cm$^{-1}$. The suppression of the scattering rate, which onsets at $T_C$ is reminiscent of that which occurs at energies below the Stoner gap in fully spin-polarized ferromagnets like CrO$_2$ \cite{singley1999}. In such cases longitudinal Stoner-like spin-flip scattering is forbidden at energies below a threshold set by the energy difference from the bottom of the minority band to the Fermi level (the Stoner gap). UGe$_2$ is not fully spin polarized but has only a small minority spin population at $E_F$ \cite{shik2001,yamagami2003}. Moreover, the behavior of properties such as the pressure dependent magnetization and mechanisms of the pairing mechanism \cite{sandeman2003} have been interpreted as a consequence of narrow peaks in the density of states, which could give an effective gap to spin flip excitations. Such a density of states is supported by band calculations \cite{shick2004}. Longitudinal fluctuations as such, which can possibly mediate exotic superconductivity, have been found by neutron scattering \cite{huxley03}. We note that conventional magnons are not expected to play a large role in a strongly unaxial compound like UGe$_2$ as their energy scales will be much higher. An effective gap to longitudinal spin-flip excitations has also been inferred through a Stoner model fit to the strength of magnetic Bragg peaks at low temperature with a gap that is on the order of the threshold in the optical scattering rate \cite{aso2005}.  It is interesting that our spectra give a strong indication of a coupling of charge to these longitudinal fluctuations that were originally proposed as a possibility to mediate superconductivity in ferromagnetic compounds.  

The observed suppression in the scattering rate is additional to that expected generically for HF compounds below their coherence temperatures.  The usual expectation is that the effective mass $m^*$ and the quasiparticle lifetime $\tau^*$ are renormalized by approximately the same factor \cite{millis1987}.  In contrast,  comparing the low temperature scattering rates and masses with their high temperature unrenormalized values we find the ratios $\frac{m^*}{m} \approx 6$ and $\frac{\tau^*}{\tau} \approx 50$, which disagree by a factor of 8. A similar analysis using instead the high frequency values gives a similar discrepancy.  The additional scattering suppression which onsets at $T_C$ implies a strong coupling between HF effects and magnetic ones with important implications for superconductivity. It is also interesting to note that as the temperature increases the energy scale of the threshold in $1/\tau(\omega)$ does not appear to close at the transition, but instead the gap `fills in' and a remnant of this suppression persists even up to 290 K. Similar effects have been observed in ferromagnetic nickel \cite{eastman1978}.  This observation of gaps which `fill in' instead of closing is a common occurrence in strongly correlated systems \cite{furukawa1999}.  

\begin{figure}[bht]
	\centering
		\includegraphics[width=5cm]{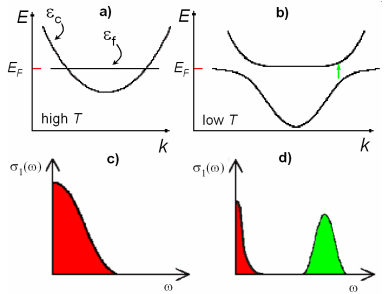}
	\caption{(Color online) a), b) Schematic band structure within the PAM of the conduction $\epsilon_c$ and the 5$f$ electrons $\epsilon_f$ at high and low temperature, respectively. Vertical arrow indicate the interband transitions. c), d) Sketch of the optical conductivity at high and low temperature.}
	\label{dispersion}
\end{figure}

The relatively large mass enhancement observed in UGe$_2$ suggests the evolution of renormalized itinerant charge carriers out of a Fermi gas coupled to a lattice of 5$f$ local orbitals. It is generally accepted that the periodic Anderson model (PAM), which describes the hybridization of a localized level with a conduction band, captures the essential physics of such systems \cite{rice1986,fulde1988,millis1987}.  In Fig. \ref{dispersion} we plot a schematic band dispersion of the PAM together with its optical conductivity at low and high temperature. At high $T$ only a dispersive conducting band crosses the Fermi level (E$_F$), whereas near E$_F$ the 5$f$ states are dispersionless. The intraband transitions give rise to a broad Drude conductivity sketched in Fig. \ref{dispersion} c). As the temperature is lowered, hybridization with the 5$f$ level splits the conduction band into two branches and is responsible for the large effective mass observed at low temperature. Excitations between the split bands create the possibility for new interband transitions (Fig. \ref{dispersion} b) and redistributes spectral weight between high and low energies as shown in Fig. \ref{dispersion} d).

Although this general picture should hold in UGe$_2$, additional features in principle are expected as a result of uranium's nominal 5$f^3$ configuration which results in several 5$f$ bands being involved in the heavy electron state.  Moreover, as has been suggested for CeCo$_{1-x}$Ir$_x$In$_5$, a distribution of hybridization gaps due to a momentum dependence of the $f-d$ coupling parameter \cite{burch2007,shim2007,weber2008} may result in multiple features in $\sigma_1(\omega)$, such as seen in Fig. \ref{s1} b) at 300 cm$^{-1}$ (37 meV) and 110 cm$^{-1}$ (13.6 meV). 

Our observations suggest an interesting interplay between spin-polarization and the HF coherent state. We believe that this is at the origin of the rather rich behaviour of the optical conductivity, resulting in two structures appearing below $T_C$ (13.6 meV and 37 meV), large quasi-particle renormalizations, and evidence for a strong coupling to the longitudinal magnetic modes which have been suggested to mediate superconductivity in this compound.

This work is supported by the Swiss National Science Foundation through Grant No. 200020-113293 and the National Center of Competence in Research (NCCR) ``Materials with Novel Electronic Properties-MaNEP'' and the European Science Foundation (ECOM-COST action P16). The authors would like to thank N. Drichko, A. Kuzmenko, F. L\'evy and F. Ronning for helpful conversations and E. Koller for technical support.

\end{document}